\documentclass{article}
\pdfoutput=1

\usepackage{arxiv}

\usepackage[utf8]{inputenc} 
\usepackage[T1]{fontenc}    
\usepackage{hyperref}       
\usepackage{url}            
\usepackage{booktabs}       
\usepackage{amsfonts}       
\usepackage{nicefrac}       
\usepackage{microtype}      
\usepackage{comment}
\usepackage{orcidlink}
\usepackage{apacite}
\PassOptionsToPackage{nosectionbib}{apacite}
\RequirePackage[]{natbib}

\usepackage[american]{babel}
\usepackage{csquotes}
\usepackage{amsmath}
\usepackage{todonotes}
\usepackage{tikz}
\usepackage{float}
\usepackage{lineno}
\usepackage{multirow}
\usepackage{adjustbox}
\usepackage{amsthm}
\usepackage{amssymb}
\usepackage{graphicx}
\usepackage[figuresright]{rotating}
\usepackage{comment}
\usepackage{tikz}
\usetikzlibrary{arrows.meta, positioning}
\usepackage{xcolor}

\title{Single-Dataset Meta-Analysis For Many-Analysts And Multiverse Studies}
\renewcommand{\shorttitle}{Single-Dataset Meta-Analysis}

\author{
    František Bartoš\orcidlink{0000-0002-0018-5573} \\
	Department of Psychological Methods \\
	University of Amsterdam             \\
	Noord-Holland, The Netherlands      \\
    \And
    Suzanne Hoogeveen\orcidlink{0000-0002-1304-8615}\\
    Department of Methodology and Statistics \\
	Utrecht University                  \\
	Utrecht, The Netherlands      \\
    \And
    Alexandra Sarafoglou\orcidlink{0000-0003-0031-685X}\\
    Department of Psychological Methods \\
	University of Amsterdam             \\
	Noord-Holland, The Netherlands      \\
    \And
    Samuel Pawel\orcidlink{0000-0003-2779-320X}\\
    Epidemiology, Biostatistics and Prevention Institute \\
    Center for Reproducible Science and Research Synthesis \\
    University of Zurich, Switzerland \\
}

\begin{document}
\maketitle

\begin{abstract}
Empirical claims often rely on one population, design, and analysis. Many-analysts, multiverse, and robustness studies expose how results can vary across plausible analytic choices. Synthesizing these results, however, is nontrivial as all results are computed from the same dataset. We introduce single-dataset meta-analysis, a weighted-likelihood approach that incorporates the information in the dataset at most once. It prevents overconfident inferences that would arise if a standard meta-analysis was applied to the data. Single-dataset meta-analysis yields meta-analytic point and interval estimates of the average effect across analytic approaches and of between-analyst heterogeneity, and can be supplied by classical and Bayesian hypothesis tests. Both the common-effect and random-effects versions of the model can be estimated by standard meta-analytic software with small input adjustments. We demonstrate the method via application to the many-analysts study on racial bias in soccer, the many-analysts study of marital status and cardiovascular disease, and the multiverse study on technology use and well-being. The results show how single-dataset meta-analysis complements the qualitative evaluation of many-analysts and multiverse studies.
\end{abstract}


\section{Introduction}

Studies are often conducted using a single specific population, manipulation, and statistical analysis. While researchers may aim to draw broader conclusions from such work, these generalizations may not be justified unless the potential variability across populations, design choices, and analytical choices is explicitly accounted for in the analysis \citep{yarkoni2022generalizability, holzmeister2024heterogeneity}. A single study is usually insufficient to capture all sources of variability.

Generalizations with respect to populations and designs can be assessed through cross-cultural studies and conceptual replications \citep{oberauer2019addressing, almaatouq2022beyond}, while robustness across similar populations and nearly identical designs is evaluated through direct replications, for instance, via large-scale multi-site replications (e.g., Many labs studies; \citealp{Klein2014ML1, Klein2018ML2, ebersole2016many, klein2022many, ebersole2020many}) and initiatives to systematically reproduce and replicate research findings (e.g., \citealp{brodeur2024reproduction}). These efforts involve collecting data from independent samples and are typically analyzed using standard meta-analytic methods. Similarly, in medicine, bottom-up collaborative approaches have gained popularity, where multiple randomized controlled trials are independently designed but follow closely aligned protocols and are jointly analyzed in a prospective meta-analysis \citep{seidler2019guide, tierney2021framework}.

Generalization with respect to statistical analyses can be assessed through many-analysts studies, multiverse studies, or robustness analyses. In many-analysts studies, different analyses performed by independent researchers on a single dataset allow researchers to assess the robustness of the findings across plausible analytic strategies \citep{silberzahn2018many}. In multiverse studies, key aspects of the analytic strategy, such as the operationalization of outcome variables or exclusion criteria, are systematically varied to create a set of plausible analytic combinations, all of which are executed and transparently reported \citep{steegen2016increasing, SimonsohnEtAl2020}. On a smaller scale, robustness analyses explore the variability of results with respect to a select few choices, for instance, different shapes of prior distributions within the Bayesian analysis framework \citep{wagenmakers2022one}.

Both many-analysts and multiverse studies help reveal the ``garden of forking paths'', where numerous analytic strategies may be explored but only those leading to significant results are reported \citep{gelman2013garden}. In addition, the many-analysts approach allows researchers to incorporate diverse perspectives and add nuance to findings by comparing results across different expert teams or theoretical frameworks, even with a small number of analysts \citep{hoogeveen2024prevalence, dongen2019multiple, bartos2025introducing}.

Although many-analysts and multiverse approaches have gained popularity in recent years (e.g., \citealp{hoogeveen2023improving, harder2020multiverse, stern2019no, wessel2020multiverse, hoogeveen2023many}), statistical methods for aggregating evidence in these projects remain underdeveloped. Some recent techniques allow for hypothesis testing within the multiverse framework \citep{mandl2024addressing, girardi2024post, coqueret2025global, SimonsohnEtAl2020}. However, these techniques require costly simulation approaches and are typically only feasible when all analyses have been developed by a single research team (as in multiverse or robustness analyses). They become prohibitively demanding in many-analysts studies. In these cases, the lead team would need to re-implement and rerun the models from all analysis teams, an effort that is often infeasible due to the substantial variation in complexity of the submitted analytic approaches. 
Moreover, beyond these practical limitations, these techniques cannot synthesize effect size estimates into an overall effect across analysis pipelines, nor can they quantify evidence for between-analyst heterogeneity. 

Synthesis of multiple effect sizes into an overall effect size is usually within the domain of meta-analytic methodology. The standard meta-analysis however assumes that effect sizes are independent and that they come from separate, non-overlapping datasets. In many-analysts studies, by contrast, all effect size estimates are derived from the same dataset, which introduces dependence among estimates and shared sources of variance. Consequently, synthesizing evidence across analytic approaches remains a major open challenge \citep{aczel2021consensus}. In the absence of such methods, visual representations are often recommended as an effective tool to present findings. And indeed, the majority of many-analysts studies rely on visualizations--such as forest plots--to summarize the evidence, and researchers often draw conclusions about the presence or absence of an effect based on descriptive statistics alone. This approach, though widely used, is ultimately unsatisfactory. Relying exclusively on descriptive approaches makes it impossible to formally test hypotheses of interest, for instance, whether there is a genuine effect of an experimental manipulation.

Nevertheless, some projects, such as \citet{gould2025same} and \citet{coretta2023multidimensional}, have applied standard meta-analysis to calculate an overall effect size across analysis teams, despite the methodological limitations posed by using a single underlying dataset. When standard meta-analysis is applied under these conditions, it may still yield a valid estimate of between-analyst heterogeneity; however, the uncertainty around the overall effect size will likely be underestimated, as the method incorrectly assumes that each analysis team contributes new, independent data.

\subsection{The Current Manuscript}

In this manuscript, we introduce a meta-analytic approach for addressing the shared use of data in many-analysts studies, multiverse analyses, and, more broadly, robustness analyses. This framework enables researchers to synthesize evidence across analysis strategies in a statistically coherent manner. It is particularly valuable when the aim is to quantify analytic variability while also drawing overall conclusions about the magnitude and presence of the effect of interest. In doing so, it supports the use of many-analysts and multiverse designs to address substantive research questions and formally test hypotheses. The approach also contributes to a more comprehensive understanding of the evidence these projects generate and complements other tools for extracting meaningful insight, such as the Subjective Evidence Evaluation Survey \citep{sarafoglou2024sees}, which captures analysis teams' beliefs about an effect's plausibility, confidence in their results, perceived robustness, and evaluations of the dataset and study design. 

In the following sections, we first describe the limitations of applying standard meta-analytic techniques to many-analysts and multiverse projects---specifically, the problem of multiple estimates derived from the same dataset resulting in overly confident conclusions. Next, we outline several desirable properties for a method that quantitatively synthesizes results when effect sizes are computed from shared data. We then introduce single-dataset meta-analysis, a two-stage procedure that meets these criteria: first estimating heterogeneity across analysis teams or analytic strategies, and then pooling the estimates via a fractional conditional likelihood. We provide both classical and Bayesian implementations that yield estimates of analytic heterogeneity, a common-effect size, and tests for the presence of heterogeneity and an overall effect. We demonstrate the method using three applied examples from the literature. For brevity, we use the term many-analysts throughout, while noting that the method applies equally to multiverse and robustness analyses.

\subsection{Availability of Data and Code}
Readers can access the data and the R code to conduct all analyses (including example applications and the reported simulation, and the creation of all figures), in our Open Science Framework (OSF) repository: \url{https://osf.io/erxf2}. 

\section{Single-Dataset Meta-Analysis}

\subsection{Limitations of Standard-Analytic Techniques in Shared-Dataset Settings}

Standard meta-analytic methods model the $K$ effect size estimates $y_k$ and their sampling variability expressed as standard errors $\text{se}_k$ (usually assumed to be known) using either the common-effect model, also referred to as the equal-effects model or fixed-effect model,
\begin{align}
    \label{eq:equal_effect}
    y_k \mid \mu \sim \text{Normal}(\mu, \text{se}_k^2) 
\end{align}
or the random-effects model,
\begin{align}
    \label{eq:random_effects}
                \theta_k \mid \mu, \tau  &\sim \text{Normal}(\mu, \tau^2)         \\
    \nonumber   y_k  \mid \theta_k     &\sim \text{Normal}(\theta_k, \text{se}_k^2). 
\end{align}

In the common-effect model, the parameter $\mu$ corresponds to the common true effect underlying all estimates. In the random-effects model, the parameters $\mu$ and $\tau$ describe a distribution of true (unobserved) effect sizes $\theta_k$ that vary across studies, with mean $\mu$ and standard deviation $\tau$, representing the between-study heterogeneity.\footnote{The common-effect model can be seen as a special case of the random-effects model with $\tau = 0$, that is, all true effect sizes being identical.} 
The formula of the standard error of the pooled meta-analytic estimate $\hat{\mu}$ is then given as,
\begin{align}
    \label{eq:var_mu}
    \text{se}(\hat{\mu}) &= \sqrt{ \frac{1}{\sum_{k = 1}^K{\frac{1}{\text{se}_k^2}}} }\\ \nonumber
    \text{se}(\hat{\mu}) &= \sqrt{ \frac{1}{\sum_{k = 1}^K{\frac{1}{\text{se}_k^2 + \tau^2}}}}
\end{align}
for the common-effect model and random-effects models, respectively \citep[see e.g.,][]{borenstein2009introduction}. Both models assume that the effect size estimates come from independent data sources. In this context, \emph{independent} means that the sampling variability of the individual estimates is uncorrelated, which is generally the case when each estimate comes from a distinct study with no overlapping participants. 

Many-analysts studies present a very similar setup: $K$ independent analysis teams are instructed to analyze the same dataset, each applying their chosen analytic approach $f_k()$, to address the research question without interacting with one another. In most cases, the teams provide effect size estimates with their standard errors, such as $\{y_k, \text{se}_k\} = f_k(\text{data})$, for instance a regression coefficient and its standard error obtained from fitting the regression model that the teams deem most plausible. We are interested in obtaining a pooled cross-analytic estimate for the effect $\mu$ and an estimate of the between-analyst heterogeneity $\tau$. Applying the standard meta-analytic methods to many-analysts' estimates, however, inherently violates the independence assumption: the analyses are based on the same dataset.

When the sampling variability of the individual estimates is dependent be (with a complete dependency for identical data), the uncertainty in the meta-analytic parameter estimates is understated, which leads to a standard error for the pooled estimate that is too small. This underestimated standard error, in turn, produces overly narrow confidence or credible intervals, inflates test statistics, and overstates statistical significance as well as posterior probabilities and Bayes factors.

\begin{figure}
    \centering
    \includegraphics[width=0.5\linewidth]{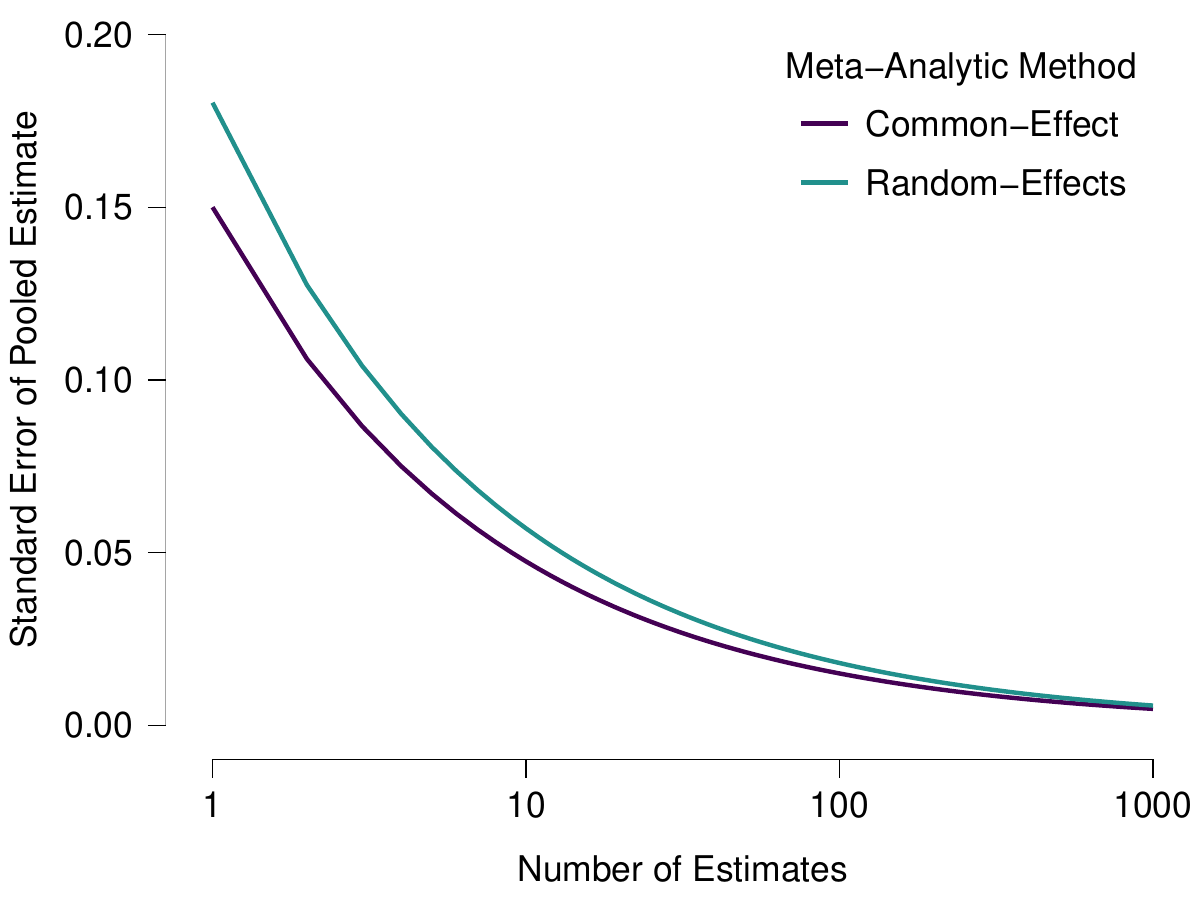}
    \caption{Standard error of the pooled meta-analytic estimate (Equation~\ref{eq:var_mu}) with increasing number of effect size estimates. All sampling variances are $\text{se}_k = 0.15$. For the random-effects model, the between-estimates heterogeneity is $\tau = 0.10$).}
    \label{fig:metaanalysis_ci}
\end{figure}

Figure~\ref{fig:metaanalysis_ci} illustrates how the expected standard error of pooled meta-analytic estimates changes as the number of effect size estimates increases. In standard meta-analysis, these estimates come from independent studies, whereas in many-analysts projects they are provided by independent analysis teams. We can distinguish between two scenarios in the meta-analytic setting; scenario 1 follows the common-effect model (purple) in which the estimates differ only due to the sampling variability in the underlying data. Scenario 2 follows the random-effects model (turquoise) in which the estimates differ due to both the sampling variability and the between-estimate heterogeneity. For simplicity, we assume equal sampling variability for all estimates ($\text{se}_k = 0.15$) in both scenarios, and we set the heterogeneity parameter to $\tau = 0.10$ in scenario 2.

The decrease of the standard error of the pooled estimate toward zero reflects the expected behavior in standard meta-analytic settings. The intuition is as follows: as the number of included studies, and thus the total sample size of the data used to estimate the pooled estimate in the common-effect model, increases, the sampling variability decreases, and the pooled estimate can be estimated with greater precision. In the random-effects model, the standard error of the pooled estimate is slightly larger because it also accounts for between-estimate heterogeneity; however, increasing the number of estimates still mitigates this additional uncertainty and reduces the standard error.

\subsection{Desired Properties in Many-Analysts Studies} \label{section:desired_properties}

In many-analysts settings, however, the intuition from standard meta-analysis does not apply. Adding more analysis teams does not increase the amount of underlying data; by definition, each team works with the same dataset. Consequently, in scenario 1, the standard error of the pooled meta-analytic estimate, which is determined solely by the sampling variability of the data, should remain constant regardless of how many analysis teams study the research question. In scenario 2, the standard error of the pooled meta-analytic estimate depends on both the sampling variability and the variability arising from differences in analytical choices (e.g., Equation~\ref{eq:var_mu}). As the number of analysis teams increases, the uncertainty in the pooled estimate due to between-analyst heterogeneity should decrease to a bound corresponding to the sampling variability of the data (i.e., scenario 1). Formally, we would expect the following formulas for the standard error of pooled cross-analytical estimate to apply, 
\begin{align}
    \label{eq:var_mu_adj}
    \text{se}(\hat{\mu}) &= \text{se}_k\\ \nonumber
    \text{se}(\hat{\mu}) &= \sqrt{\text{se}_k^2 + \frac{\tau^2}{K}},
\end{align}
had all analyses yielded an equal standard error $\text{se}_k$. Imagine an extreme case in which infinitely many analysis teams participate in a many-analysts project, all conducting an identical analysis and producing the same effect size estimate. Under the standard meta-analytic method, adding these duplicate analyses would eliminate uncertainty: because the method assumes each estimate comes from independent data, confidence or credible intervals of the pooled estimate would shrink to the point estimate, $p$-values would decrease to zero (unless $\mathcal{H}_0$ is true), and the Bayes factors would decrease to zero (under $\mathcal{H}_0$) or increase to infinity (under $\mathcal{H}_1$). In reality, however, additional analyses in many-analysts projects contribute no new data. A series of identical analyses should not reduce or eliminate uncertainty, as nothing new about the underlying effect is learned from one analysis to the next. The pooled meta-analytic effect estimate should be identical to that of any analysis team under the common-effect model. This would also be true under the random-effects model since the between-analyst heterogeneity is zero in this extreme case.

Now imagine another extreme case in which infinitely many analysis teams participate in a many-analysts project, each conducting a different analysis and producing a different effect size estimate with the same standard error. As before, increasing the number of teams to infinity in a common-effect model should not reduce the standard error of the pooled effect size estimate; instead, the standard error of the pooled effect size estimate should remain equal to the standard error from any single analysis as additional estimates are included. In the random-effects model, however, as the number of teams increases to infinity, the standard error of the pooled effect size estimate should decrease to a bound corresponding to the common-effect model since adding more analysis teams provides additional information about between-analyst heterogeneity (which also contributes to the standard error of the pooled estimate, cf. Equation~\ref{eq:var_mu_adj}).

\subsection{Single-Dataset Meta-Analysis as Special Case of a Sample Overlap}

The many-analysts setting is similar to the sample overlap problem sometimes encountered in meta-analyses of observation studies \citep{bom2020generalized}. For instance, different studies report an association between gross domestic product (GDP) and the degree of industrialization in the world in different periods. If those periods overlap, the meta-analysis needs to adjust for the underlying data-overlap between studies, otherwise it risks underestimating standard errors of the pooled estimate \citep{bom2020generalized}. 

The degree of overlap in many-analysts studies is, however, beyond the settings for which the sample overlap adjustment was originally developed; all analysis teams are given the same ``input'' dataset, which suggests perfect overlap in principle. While individual data cleaning steps and transformations can slightly reduce correlations among outcome variables, the actual extent of overlap is difficult, if not impossible, to determine reliably. Moreover, since all analysis teams start from identical data, they provide no unique information about the underlying effect by design, only about the degree of analytic variability. 

To develop a method suited for projects based on a single dataset, we therefore assume complete data overlap, treating all effect size estimates as fully dependent. Under this assumption, a meta-analytic approach to many-analysts studies reduces to a meta-analysis with $K$ perfectly correlated effect size estimates, resulting in a singular model specification (a 100\% sample overlap with the same response variable implies the between-estimate correlation of one; e.g., \citealp{bom2020generalized}). Importantly, the assumption of complete overlap serves as a conservative safeguard: it yields the upper bound of the pooled estimate's standard error and thus prevents its underestimation.

\subsection{Single-Dataset Meta-Analysis}

In standard meta-analysis, the likelihood of the observed effect size estimate given the model parameters and its sampling variability contains the information from the data. For the single-dataset case, we address the issue of shared data dependency by proposing a simple adjustment to the standard model: down-weighting the contribution of each estimate's likelihood with respect to the pooled effect proportionally to the number of analyses conducted on the same dataset. In this way, the information from the underlying data is incorporated into the pooled effect size estimate at most once. 

Practically, this is achieved by modifying the likelihood functions of both the common-effect and random-effects models, raising the likelihood of each estimate to the power of a weight $w_k$,
\begin{align}
    \label{eq:equal_effect_likelihood_adjusted}
    p(\text{data} \mid \mu) = \prod_{k=1}^K p(y_k, \text{se}_k \mid \mu)^{w_k},   
\end{align}
for the estimation of the pooled effect in the common-effect model and
\begin{align}
    \label{eq:random_effects_likelihood_adjusted}
    p(\text{data} \mid \theta_{1, \dots, K}, \mu, \tau) = \prod_{k=1}^K p(y_k, \text{se}_k \mid \theta_k)^{w_k} \prod_{k=1}^K p(\theta_k \mid \mu, \tau),
\end{align}
for the estimation of the pooled effect in the random-effects model, such that all weights sum to one 
\begin{align}
    \nonumber   \sum_{k=1}^K w_k = 1.
\end{align}

See Appendix~\ref{app:AppendixA} for technical details. This proposal ensures that when there is complete sample overlap, information about the sampling variability with respect to the pooled effect is incorporated only once, as each team contributes only a fraction to the overall likelihood. The likelihood in the single-dataset meta-analysis corresponds to the weighted geometric mean of the likelihoods, or equivalently, the weighted arithmetic mean of the log-likelihoods contributed by each team. Appendix~\ref{app:AppendixB} illustrates with a simulation how this adjustment does not lead to the standard error inflation associated with standard meta-analytic methods and follows the desired properties outlined above.

Our proposal is inspired by the fractional Bayes factor approach \citep[e.g.,][]{ohagan1995fractional, gu2018approximated} where the likelihood is raised to a fractional power $g$ to construct a proper, data-informed prior from an otherwise improper prior. The remaining fraction of the likelihood, raised to the power $1-g$, is then used for model comparison. This separation prevents the problem of ``double use'' of the same data in both prior construction and model updating \citep[see e.g.,][for other fractional approaches]{bhattacharya2019bayesian}.

A simple choice for setting $w_k$ is to assign $w_k = 1/K$, so that each analysis team contributes an equal share of information and contributes equally to the estimation of the common-effect size. However, more elaborate weighting schemes might be incorporated to account, for instance, for differences in the quality of the submitted analyses or the analysts' expertise. Another option is accounting for the fact that some teams submit multiple estimates \citep[e.g.,][]{hoogeveen2023many, coretta2023multidimensional} by dividing their allocated weight among all submitted estimates. This latter approach may be particularly appealing since only a single main estimate per team is typically presented in the main manuscript in many-analysts studies, while additional estimates are relegated to appendices or omitted altogether.

\subsection{Classical Model Estimation}

Under the single-dataset common-effect model (Equation~\eqref{eq:equal_effect_likelihood_adjusted}), maximizing the fractional likelihood with respect to $\mu$ corresponds to ordinary meta-analytic estimation that uses standard errors divided by the weights. This means that if equal weights are assigned ($w_k = 1/K$), the same meta-analytic point estimate (but not the standard error) is obtained as with a standard common-effect meta-analysis, since the weights become a multiplicative constant that cancels out. The standard error of the meta-analytic estimate differs between the single-dataset and the standard common-effect meta-analysis, with the standard meta-analysis underestimating the uncertainty.

Under the single-dataset random-effects model (Equation~\eqref{eq:random_effects_likelihood_adjusted}), $\tau$ and $\mu$ need to be estimated in two separate steps. That is because the likelihood with respect to between-analyst heterogeneity $\tau$ does not require the fractional adjustment---estimates from each team contribute unique information to $\tau$; however, the likelihood with respect to the common true effect $\mu$ requires the fractional adjustment---estimates from the analysis teams contribute redundant information with respect to $\mu$. To address this, we propose first estimating $\tau$ with a standard random-effects model and then plugging the $\tau$ estimate into a second-stage model with the standard errors multiplied by weights to obtain the pooled cross-analytical estimate and its standard error. Although this procedure seems ad hoc, it is similar to the standard classical meta-analytic approach which estimates the pooled effect size conditional on the heterogeneity estimate (which is typically also estimated using a different approach than the mean effect size, e.g., restricted maximum likelihood or method of moments). Unlike the common-effect model, the pooled estimate from a single-dataset random-effects meta-analysis does not correspond to the pooled estimate from a standard random-effects meta-analysis, even when equal weights are specified. However, as the number of analysts increases, the pooled estimate (and its uncertainty) from the single-dataset random-effects meta-analysis approaches the pooled estimate from the single-dataset common-effect meta-analysis. This is a consequence of the decreasing weights ``blowing up'' the standard errors; the heterogeneity stops influencing the weighted average as it is much smaller than the inflated standard error.

The single-dataset meta-analysis can be performed using the \texttt{metafor} \texttt{R} package \citep{metafor}. The common-effect model can be estimated by dividing the sampling variances by weights $w$, conveniently set as $w = 1/K$, 
\begin{verbatim}
  library("metafor")
  common <- rma(yi = y, vi = (se^2)/w, method = "FE")
\end{verbatim}
The random-effects model requires a two-stage procedure. First, the between-analyst heterogeneity is estimated using the standard meta-analytic model. Second, the pooled cross-analytical estimate is obtained by dividing the sampling variances by $w_k$ and using a fixed between-analyst heterogeneity obtained from the previous stage.
\begin{verbatim}
  random1 <- rma(yi = y, vi = se^2)
  random2 <- rma(yi = y, vi = (se^2)/w, tau2 = random1$tau2)
\end{verbatim}

\subsection{Bayesian Model Estimation}

The Bayesian version of the procedure closely mirrors the standard approach. For the single-dataset common-effect meta-analysis, we can use the \texttt{RoBMA} \texttt{R} package \citep{RoBMA} to estimate a common-effect model with the sampling variances divided by weights $w$,
\begin{verbatim}
  library("RoBMA")
  common <- NoBMA(y = y, v = (se^2)/w,
                  priors_effect = prior(...),
                  priors_heterogeneity = NULL)
\end{verbatim}
where `priors\_effect' parameter sets a prior distribution for the effect size, and setting `priors\_heterogeneity = NULL' specifies a common-effect model.
The random-effects model, again, requires a two-stage procedure, where the `random1' model estimates the between-analyst heterogeneity under a prior distribution specified via the `priors\_heterogeneity' argument, which is then fixed in the second stage via the same argument 
\begin{verbatim}
  random1 <- NoBMA(y = y, v = se^2,
                   priors_effect = prior(...),
                   priors_heterogeneity = prior(...),
                   priors_heterogeneity_null = NULL)
  posterior_tau <- extract_posterior(random1, "tau")
  random2 <- NoBMA(y = y, v = (se^2)/w,
                   priors_effect = prior(...),
                   priors_heterogeneity = prior("spike", list(median(posterior_tau))),
                   priors_heterogeneity_null = NULL, algorithm = "ss")
\end{verbatim}
The `algorithm = "ss"' argument is necessary for using the product-space algorithm implemented in the package that allows specifying a point prior distribution equal to the $\tau$ estimate from the first stage, ensuring the correct $\tau$-value is used in the second stage.

The `prior(...)` function can be used to define many types of prior distributions (e.g., normal, Student-t, gamma, beta; see `?prior` for details). Researchers can adopt default prior distributions from existing statistical tests (e.g., \citealp{rouder2009bayesian}), specify a wide range of informed prior distributions (e.g., \citealp{gronau2020informed}, or perform prior elicitation \citealp{johnson2010methods, chaloner1996elicitation, ohagan2006uncertain, mikkola2021prior}), in the case of a many analysts study re-examining a previous finding, use the original estimate and obtain a replication Bayes factor \citep{ly2019replication}, or specify wide weakly-informative prior distributions if they are interested in parameter estimation only \citep{rover2021weakly}. If prior information is lacking and researchers wish to perform a hypothesis test, normal prior distribution with a standard deviation set to a unit information \texttt{UI} can be specified for the pooled effect size as described in Chapter 2.4 in \cite{spiegelhalter2004bayesian} and in Chapter 1 in \cite{grieve2022hybrid} and a half-normal prior distribution with standard deviation equal to one half of \texttt{UI} can be specified for the between-analyst heterogeneity under alternative hypotheses. Unit information prior distributions allow us to specify prior distributions across a wide range of effect size measures \citep[also see][]{mulder2024bayesian, rover2021weakly}. They provide weak regularization for the estimates and conservative hypothesis tests. By default, the null hypotheses are specified as spikes at 0 (i.e., all prior mass is concentrated at a single point; however, different null hypotheses, such as peri-null distributions, can also be specified \citealp{ly2022bayes}, or set to `NULL` for an estimation-only approach; see `?NoBMA` for details).\footnote{Alternatively, we could use the previously obtained pooled effect size estimate and its standard error from the classical approach and combine it with a prior distribution using the approximate normal likelihood \citep{tsou1995robust, royall1997statistical} to obtain the posterior distribution and the Bayes factor \citep[i.e., Bayesian z-test;][]{berger1987testing, bartos2022fast}.}

\section{Examples}

In this section, we apply our methodology to two published many-analysts studies \cite{silberzahn2018many, kowall2025marital} and to a multiverse analysis \cite{orben2019social}. We report the results of both the classical and Bayesian single-dataset random-effects meta-analyses. As this is an exploratory study, $p$-values are not dichotomized but interpreted as quantitative measures of evidence against the null hypothesis, i.e., following a Fisherian rather than the Null Hypothesis Significance Testing (NHST) framework which is a widely used but incoherent blend of Fisherian and Neyman-Pearson inference \citealp{Gigerenzer2004, bland2015introduction, Perezgonzalez2015, goodman2016aligning, Greenland2023}). For the Bayesian analyses, we follow a Jeffreys approach and report posterior estimates and credible intervals conditional on the alternative hypothesis and test for the presence vs. absence of the effect/heterogeneity via Bayes factors \citep{jeffreys1935some, jeffreys1961theory, kass1995bayes} based on unit information priors as outlined in the previous paragraph. 

\subsection{Silberzahn et al.: Racial Bias in Soccer} 

The first example we present is the seminal many-analysts study by \cite{silberzahn2018many}, which investigated racial bias in soccer. The central research question asked whether professional referees were more likely to give red cards to players with darker skin tones compared to those with lighter skin tones. The dataset comprised archival records for $2,053$ players in the top divisions of England, Germany, France, and Spain during the 2012--2013 season, linked to $3,147$ referees across the players' careers. In total, it included $146,028$ player–referee interactions, along with demographic information, match statistics, referee decisions (including yellow and red cards), and player skin tone ratings, which were coded from photographs on a five-point scale ranging from very light to very dark and then rescaled to values between 0 (very light) and 1 (very dark).

\begin{figure}
    \centering
    \includegraphics[width=0.65\linewidth]{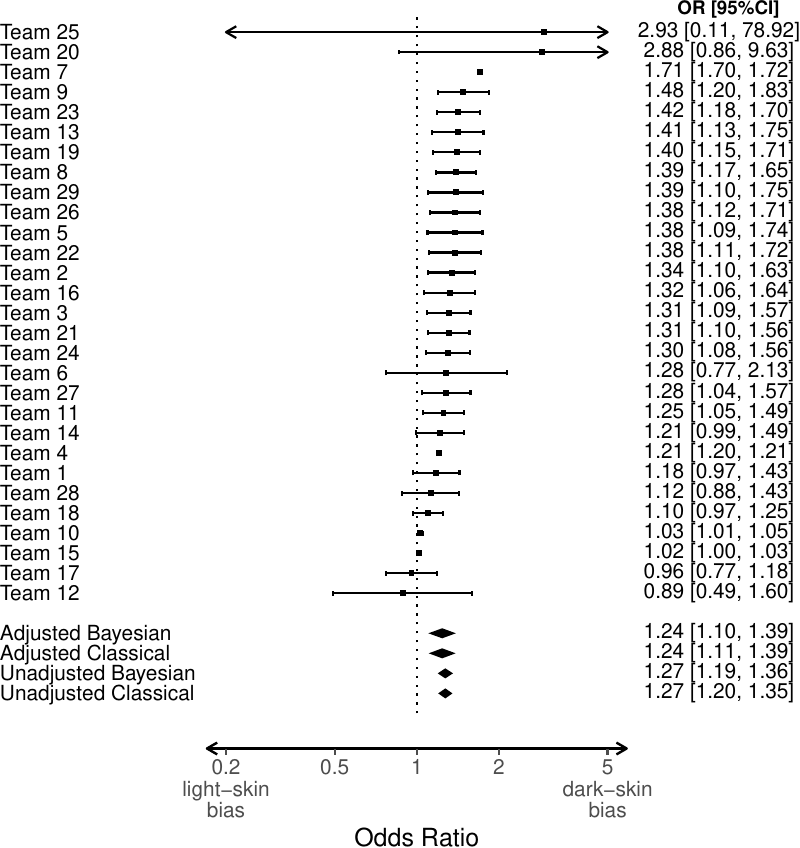}
    \caption{Forest plot of the main results from the many-analysts project on racial bias in soccer conducted by \cite{silberzahn2018many}. Supporting the original authors' conclusions, our reanalysis indicates strong evidence for a bias against darker-skinned players (\(\mu_\text{OR} =\) 1.24 {[}1.10, 1.39{]}; \(\text{BF}_{10} = 24.0\)), as well as strong evidence supporting the presence of heterogeneity (\(\tau_\text{logOR} = \) 0.13 {[}0.10, 0.19{]}; \(\text{BF}_{10} > 10^{308}\)).}
    \label{fig:silberzahn_forrest}
\end{figure}

In total, 29 independent analysis teams from the social sciences (e.g., psychology, economics, sociology, and management) and statistics were recruited. Reported associations, expressed as odds ratios, ranged from $0.89$ to $2.93$, with a median of $1.31$ (see Figure\ref{fig:silberzahn_forrest}). A majority of teams (69\%) found a statistically significant positive association, suggesting bias against darker-skinned players, while 31\% did not observe a statistically significant relationship. The authors concluded that although there was a general trend toward a positive association, the heterogeneity of results across teams demonstrated how subjective yet defensible analytic decisions can substantially shape research findings.

The classical single-dataset random-effects meta-analysis of these data finds strong evidence for dark-skin bias, \(\hat{\mu}_\text{OR} =\) 1.24 (95\% CI from 1.11 to 1.39; \(p =\) 0.0002) along with strong evidence for small but non-negligible between-analyst heterogeneity, \(\hat{\tau}_\text{logOR} = \) 0.13 (95\% CI from 0.08 to 0.17; \(p <\) 0.0001), supporting the conclusions of the original authors with statistically valid inferences. While the original authors noted a general trend toward a positive association, we can confirm this conclusion and find strong evidence against the null hypothesis of no association. 

The Bayesian single-dataset random-effects meta-analysis finds strong evidence for the presence of bias against darker-skinned players, with a Bayes factor of $\text{BF}_{10} = 24.0$ favoring the alternative model ($\mu_\text{logOR} \sim \text{N}(0, 2)$\footnote{For log odds ratios and relative risks, the standard deviation of the prior based on a unit information UI corresponds to 2 by convention \citep{spiegelhalter2004bayesian}.}) over the null model ($\mu_\text{logOR} = 0$). The pooled cross-analytical effect \(\mu_\text{OR}\) has a posterior median of 1.24 (95\% CrI from 1.10 to 1.39), virtually identical to the classical result. Consistent with the subjective assessment of the original authors, we also find evidence for heterogeneity across analyses, with a Bayes factor larger than \(10^{308}\) supporting a random-effects model ($\tau_\text{logOR} \sim \text{N}_+(0, 1)$) over a common-effect model ($\tau_\text{logOR} = 0$). The heterogeneity parameter \(\tau_\text{logOR}\) has a posterior median of 0.13 (95\% CrI from 0.10 to 0.19), which again very closely mirrors the classical result.

In contrast, the standard meta-analysis would deflate the standard error of the pooled estimate and inflate the evidence for the presence of a positive association: \(\hat{\mu}_\text{OR} =\) 1.27 (95\% CI from 1.20 to 1.35; \(p < 0.0001\)) using the classical approach and \(\mu_\text{OR} =\) 1.27 (95\% CrI from 1.19 to 1.36; \(\text{BF}_{10} >\) 1000) using the Bayesian approach.

\subsection{Kowall et al.: Marital Status And Cardiovascular Disease}

The second example is a recently published many-analysts study by \cite{kowall2025marital}, which investigated whether marital status influences the incidence of cardiovascular disease. The analysis teams were instructed to investigate whether unmarried individuals had a higher risk of developing cardiovascular disease compared to married individuals. The dataset was drawn from the Survey of Health, Ageing and Retirement in Europe (SHARE), a large multinational longitudinal panel study of $140,000$ individuals aged 50 years and older from 28 European countries and Israel. SHARE began in 2004 and is harmonized across biennial follow-up waves. For this study, the teams were instructed to use only waves 1–7 (i.e., data from 2004--2017). Cardiovascular events, such as heart attack or stroke, were compared between individuals who had never married and those who lived with a partner.

\begin{figure}
    \centering
    \includegraphics[width=0.65\linewidth]{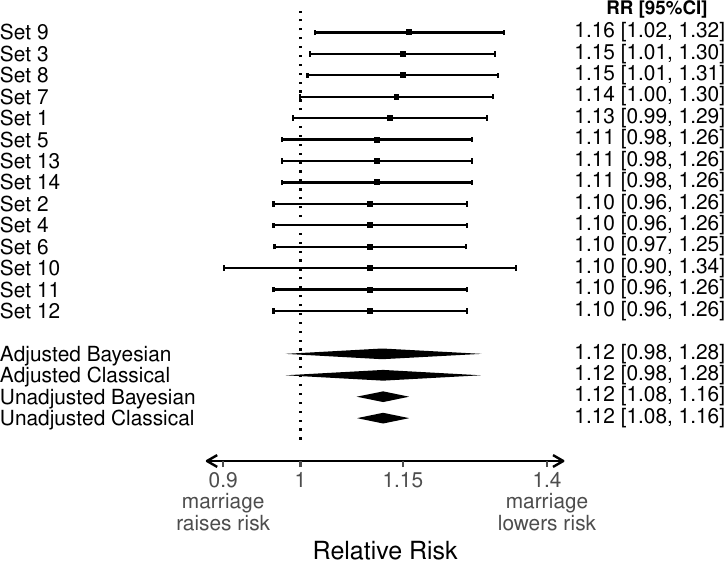}
    \caption{Forest plot of the results from adjustments sets analysis in the many-analysts project on the relationship between marital status and cardiovascular disease conducted by \cite{kowall2025marital}. Our reanalysis provides moderate evidence in favor of the null hypothesis that marital status is not associated with cardiovascular events (\(\mu_\text{RR} =\) 1.12 {[}0.98, 1.27{]}; \(\text{BF}_{01} = 7.29\)), alongside strong evidence for the absence of heterogeneity (\(\tau_\text{logRR} = \) 0.01 {[}0.00, 0.05{]}; \(\text{BF}_{01} = 47.8\)).}
    \label{fig:kowall_forrest}
\end{figure}

In total, 16 independent analysis teams, primarily from epidemiology, medicine, and statistics, were recruited. For the main analysis, the teams provided different association measures -- odds ratios, hazard ratios, and relative risks. While all pertaining to the relation between marital status and cardiovascular disease, these associations are not directly comparable and answer slightly different questions. In an additional analysis, the authors identified the different adjustment sets (i.e., covariates) used by the teams. In order to investigate sources of variability, the authors then estimated the relative risks in a longitudinal analysis using these different adjustment sets. Here, we focus on the aggregated associations from this additional analysis, which are on the same scale (relative risk) and range from 1.10 to 1.16 (see Figure~\ref{fig:kowall_forrest}).

The classical single-dataset random-effects meta-analysis extends the assessment of the original authors. While the authors did not draw conclusions on the overall association across the different adjustments sets specifically, their conclusion of no clear overall association in longitudinal analyses is corroborated by our analysis; we find weak evidence against the null hypothesis of no association, $p = $ 0.097, with the the pooled meta-analytic estimate \(\hat{\mu}_\text{RR}\) of 1.12 (95\% CI from 0.98 to 1.28) and no evidence for between-analyst heterogeneity \(\hat{\tau}_\text{logRR} = \) 0.00 (95\% CI from 0.00 to 0.00; \(p =\) 1; the Q-profile method produces a CI consisting of a point due to virtually no observed heterogeneity).

Similarly, the Bayesian single-dataset random-effects meta-analysis provides moderate evidence against a relation between marital status and cardiovascular disease, with the data being 7.29 times more likely under the null hypothesis model ($\mu_\text{logRR} = 0$) versus the alternative hypothesis model ($\mu_\text{logRR} \sim \text{N}(0, 2)$). The pooled cross-analytical estimate \(\mu_\text{RR}\) has a posterior median of 1.12 (95\% CrI from 0.98 to 1.28). In addition, assessment of the heterogeneity quantifies the authors' subjective evaluation of `only a small effect' across adjustments sets, such that we find strong evidence \emph{against} heterogeneity across analyses, with a Bayes factor of 47.3 supporting a common-effect (\(\tau_\text{logRR} = 0\)) over a random-effects model ($\tau_\text{logRR} \sim \text{N}_+(0, 1)$). The heterogeneity parameter \(\tau_\text{logRR}\) has a posterior median of 0.01 (95\% CrI from 0.00 to 0.05). Together, the results indicate that any association between marital status and cardiovascular disease is likely small, with point estimates around 1.1 and confidence/credible intervals spanning values below one. Importantly, this pattern is consistent across adjustment sets.

Although the point estimates across single-dataset and standard approaches are virtually equal, the deflated standard errors in the standard meta-analyses increase the evidence against the null hypothesis of no association. Notably, using the Bayesian approach, the standard meta-analysis would indicate substantially stronger evidence for an overall association to the extent that it would suggest the data to be 120456 times \emph{more} likely under the effect model than under the null model. This finding would meaningfully shift the interpretation toward stronger support for the conclusion that unmarried people are at higher risk of developing cardiovascular disease. 

\subsection{Orben et al.: Technology Use And Well-Being}

The third example we present is a multiverse analysis published by \cite{orben2019social}. The central research question asked whether greater digital technology use was associated with lower well-being among adolescents. For this purpose, the authors used, among others, data from the UK Millennium Cohort Study, a large-scale, nationally representative longitudinal study tracking over $11,000$ children born in 2000–2002 in the United Kingdom. Data cover information on social, behavioral, and health factors, including multiple measures of screen-based digital technology use (e.g., smartphone, computer, television) and various self-reported well-being indicators such as depressive symptoms, happiness, and social connectedness.

The authors applied a multiverse analysis to systematically estimate results across thousands of plausible model specifications. In total, $20,776$ distinct specifications were analyzed. Reported effect size estimates, expressed as standardized regression coefficients (beta), were generally small and ranged from $-0.266$ to $0.165$, with a median of $-0.032$. The majority of specifications showed negative associations between technology use (including social media, TV, games) and well-being. However, the authors concluded that while the overall trend pointed toward a negative association, the effects were small in magnitude. They further noted that the apparent negative association was primarily driven by models without control variables (e.g., models that did not adjust for sociodemographic and family background). Among the specifications that included control variables, 37\% (i.e., 3668 out of 10002) produced effect sizes overlapping with zero, 39\% were negative, and 24\% were positive. Based on this pattern, the authors emphasized the wide variation across specifications.

\begin{figure}
    \centering
    \includegraphics[width=0.65\linewidth]{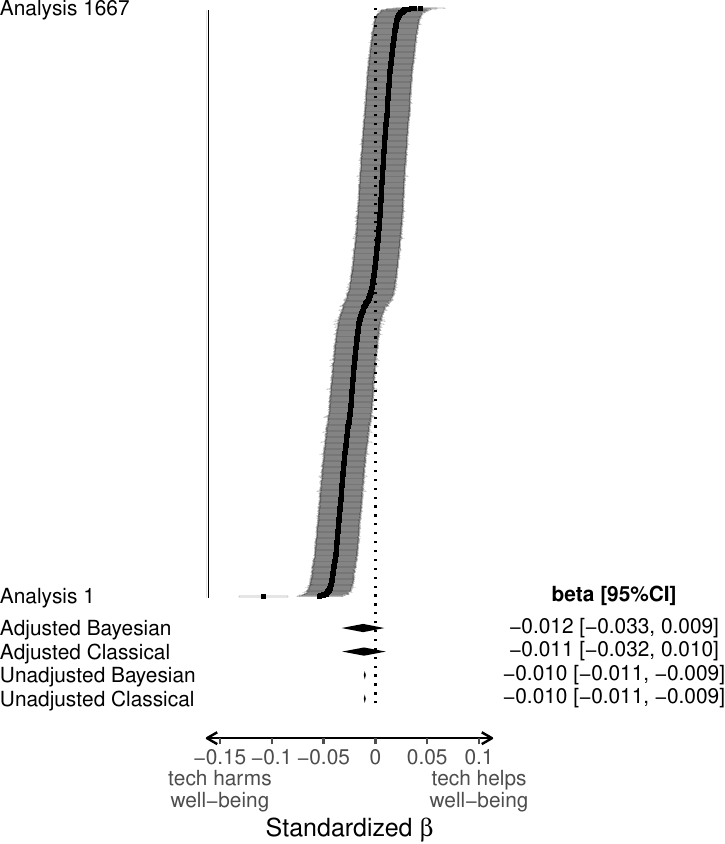}
    \caption{Forest plot of the main results from the multiverse analysis on the relationship between digital technology use and well-being in adolescents conducted by \cite{orben2019social}. Consistent with the findings of the original authors, our reanalysis of data from the UK Millennium Cohort Study, focusing on model specifications that included control variables, provides strong evidence for the null hypothesis of no substantial relationship between digital technology use and well-being (\(\mu =\) -0.01 {[}-0.03, 0.01{]}; \(\text{BF}_{01} = 49.5\)), as well as strong evidence supporting the presence of heterogeneity (\(\tau =\) 0.017 {[}0.016, 0.018{]}; \(\text{BF}_{10} > 10^{308}\)).}
    \label{fig:orben_forrest}
\end{figure}

Our reanalysis focused on data from the Millennium Cohort Study, model specifications that included control variables and used overall digital technology use as the independent variable (1667 paths). In line with the original authors, who noted that no clear overall association was observed for these specifications, the classical single-dataset random-effects meta-analysis provides no evidence for an association between digital technology and well-being $\hat{\mu} = -0.01$ (95\% CI from -0.03 to 0.01; $p=.306$). At the same time, also consistent with the original interpretation, there is strong evidence for between-analysis heterogeneity, yet the extent of heterogeneity is estimated to be small: $\hat{\tau} = 0.017$ (95\% CI from 0.016 to 0.019; p < 0.0001). 

The Bayesian single-dataset random-effects meta-analysis similarly indicates strong evidence against the association between digital technology use and well-being, with a Bayes factor of 49.7 favoring the null model ($\mu_\text{beta} = 0$) over the alternative model ($\mu_\text{beta} \sim \text{N}(0, 0.87)$, where 0.87 is the implied unit information-based scale; Eq.~6 in \citealp{rover2021weakly}). The pooled cross-analytical effect \(\mu_\text{beta}\) has a posterior median of -0.01 (95\% CrI from -0.03 to 0.01). In addition, we find strong evidence for heterogeneity across specifications, with a Bayes factor \(>10^{308}\) supporting a random-effects model ($\tau_\text{beta} \sim \text{N}_+(0, 0.43)$) over the common-effect model ($\tau_\text{beta} = 0$). The heterogeneity parameter \(\tau\) has a posterior median of 0.017 (95\% CrI from 0.016 to 0.018), virtually identical with the classical estimate and confidence interval.

Again, the inappropriate deflation of the standard error in the standard model would erroneously result in an overly precise estimate for the pooled effect size --\(\mu =\) -0.010 (95\% CrI from -0.011 to -0.009)-- and hence very strong evidence in favor of an overall effect: \(\text{BF}_{10} = 2.3\times10^{78}\) for the standard Bayesian analysis; \(p<.0001\) for the classical analysis. This example shows that especially in cases with a large number of estimates derived from the same data (e.g., large many-analysts projects or multiverses), without adjustment, the evidence for an effect can easily be greatly overstated.

\section{Discussion}

Empirical claims often rest on a single population, design, and analysis. Many-analysts studies, multiverse studies, and robustness analyses have recently become valuable tools for exposing the consequences of analytic flexibility. However, the lack of statistical tools for summarizing the results across different analyses has limited the ability to communicate a single encompassing summary estimate. We proposed single-dataset meta-analysis, a simple way to synthesize results from multiple analyses based on a single dataset. 

The single-dataset meta-analysis downweighs the likelihood contribution of each estimate so that the total information from all analyses does not exceed the likelihood contribution had only a single analysis been conducted. In contrast to standard meta-analysis, single-dataset meta-analysis does not overstate the evidence and does not underestimate the sampling variability of the data with the inclusion of more analyses performed on the same dataset. Similar to a standard meta-analysis, single-dataset meta-analysis provides easy to interpret estimates of the overall effect and the between-analysis heterogeneity, both of which can be accompanied by either classical or Bayesian hypothesis tests.

We demonstrated two-stage classical and Bayesian implementations of a common-effect and random-effects version of the single-dataset meta-analysis. The random-effects version relies on two-stage estimation. The first stage estimates the between-analyst heterogeneity with a standard random-effects meta-analysis, and the second stage modifies the standard random-effects meta-analysis by incorporating the heterogeneity estimate from the first stage and adjusting the sampling variances of the estimates by likelihood weights. 

By default, we specify equal likelihood weights across estimates, however, different weighting schemes can be used to adjust for the quality of the analysis or to deal with multiple estimates provided by the same analysis team. The procedure assumes that the analysis teams perform the analyses independently, that is, the only dependence between the estimates arises from the shared dataset. As such, the procedure can be seen as an extreme case of a sample overlap adjustment \citep{bom2020generalized} applied to a meta-analysis of a single study.

A notable practical advantage of the current approach is that it does not require the many-analysts lead team to reproduce the individual teams' analyses or to access the underlying raw data. Instead, it relies solely on the summary estimates provided by the teams. This feature makes the method uniquely feasible and scalable compared to alternative proposed approaches that require reproducing and extending individual analyses \citep[e.g.,][]{mandl2024addressing, girardi2024post, coqueret2025global, SimonsohnEtAl2020}. 

Importantly, application of a single-dataset meta-analysis is only warranted when analysis teams target a common estimand. In other words, if different teams interpret the analysis task differently, their estimates reflect distinct effects or associations that cannot meaningfully be pooled \citep{rohrer2025can}. Consequently, such estimates should not be combined in a single analysis, an issue that applies to any evidence synthesis technique. Although this assumption may be difficult to verify in practice, severe violations could manifest as a multi-modal distribution of effect size estimates. In practice, multi-analyst projects need to give participants a sufficiently precise question that leaves no ambiguity about the estimand.

The single-dataset meta-analysis effectively assumes that the analyzed estimates are a representative sample of effect sizes describing the phenomenon of interest; the observed effect sizes represent a random, unbiased sample from the broader population of plausible analyses that reflect the phenomenon. This assumption is generally easier to justify in many-analysts studies than in multiverse studies. In the former, the teams select analysis pipelines that are both plausible and typical within the research field, yet collectively span a broad range of analytic variability (but see e.g., several teams in \citealp{silberzahn2018many}, ignoring the multilevel structure of the data). By contrast, it might be more difficult to justify the representativeness assumption in multiverse studies, which may include many implausible analyses (\citealp{auspurg2025robustness}; but see e.g., \citealp{sarma2024milliways}, for possible solutions), or in robustness analyses that vary only a limited subset of plausible specifications.

Taken together, single-dataset meta-analysis offers a conservative, practical, and transparent synthesis when multiple defensible analyses are applied to the same dataset. By counting the sampling information once, it preserves appropriate uncertainty, yields interpretable estimates of average effect and between-analyst heterogeneity, and enables hypothesis tests without spurious precision.

\section*{Declarations}

Conceptualization:
FB; 
Data curation:
FB, SH, AS; 
Formal analysis:
SH; 
Funding acquisition:
AS; 
Investigation:
FB, SH, AS, SP; 
Methodology:
FB, SP; 
Visualization:
FB, SH; 
Writing – original draft:
FB, SH, AS, SP; 
Writing – review \& editing:
FB, SH, AS, SP

\subsection*{Funding}
A.S. was supported by a 2024 Ammodo Science Award and by a Veni grant from the Netherlands Organisation for Scientific Research (NWO; VI.Veni.241G.007).

\subsection*{Competing interests}
The authors declare that there were no conflicts of interest with respect to the authorship or the publication of this article.

\subsection*{Ethics approval}
This is a non-empirical study and therefore did not require ethical approval.


\renewcommand{\thesection}{A\arabic{section}}
\renewcommand{\thesubsection}{A\arabic{section}.\arabic{subsection}}
\renewcommand{\thefigure}{A\arabic{figure}}
\setcounter{section}{0}
\setcounter{figure}{0} 

\section{Appendix}

\subsection{Repeated Use of the Same Data}
\label{app:AppendixA}

When applied to many-analysts or multiverse studies, standard meta-analytic methods clearly violate the desired behavior with respect to the uncertainty in the pooled cross-analytical effect size estimate. To motivate our proposed solution, we examined the likelihoods of standard common-effect and random-effects meta-analytic models when applied to the extreme scenarios outlined in section~\ref{section:desired_properties}, that is, all teams producing an identical analysis: $f_k = f_1$ for all $k = 1, \dots, K$.

The likelihood of the common-effect model from Equation~\eqref{eq:equal_effect}
\begin{align}
    \label{eq:equal_effect_likelihood}
    p(\text{data} \mid \mu) = \prod_{k=1}^K p(y_k, \text{se}_k \mid \mu)    
\end{align}
simplifies to 
\begin{align}
    \label{eq:equal_effect_likelihood_simple}
    p(\text{data} \mid \mu) = \prod_{k=1}^K p(f_k(x) \mid \mu)  = p(f_1(x) \mid \mu)^K
\end{align}
under the assumption of identical analyses ($f_k = f_1$ for all $k = 1  , \dots, K$). The right side of Equation~\ref{eq:equal_effect_likelihood_simple} highlights the repeated use of the same data, which violates the assumption of independence of the effect size estimates. In fact, the standard meta-analytic model uses the same data exactly $K$ times.

The likelihood of the random-effects model from Equation~\eqref{eq:random_effects}\footnote{Notice that we do not marginalize the true effects $\theta_1, \dots, \theta_K$ since we want to keep independence in $p(\theta_1, \dots, \theta_K \mid \mu, \tau)$ later.} 
\begin{align}
    \label{eq:random_effects_likelihood}
    p(\text{data} \mid \theta_{1, \dots, K}, \mu, \tau) = \prod_{k=1}^K p(y_k, \text{se}_k \mid \theta_k) \prod_{k=1}^K p(\theta_k \mid \mu, \tau)
\end{align}
simplifies to 
\begin{align}
    \label{eq:random_effects_likelihood_simple}
    p(\text{data} \mid \theta_{1, \dots, K}, \mu, \tau) = \prod_{k=1}^K p(f_k(x) \mid \theta_k) \prod_{k=1}^K p(\theta_k \mid \mu, \tau) = p(f_1(x) \mid \theta_k)^K  \prod_{k=1}^K  p(\theta_k \mid \mu, \tau).
\end{align}
under the assumption of identical analyses ($f_k = f_1$ for all $k = 1  , \dots, K$). The first part of the right-hand side in Equation~\ref{eq:random_effects_likelihood_simple} again highlights repeated use of the same data (i.e., the data are incorporated into the analysis multiple times unless $K = 1$).

\subsection{Demonstration of the Desired Properties of Single-Dataset Meta-Analysis}
\label{app:AppendixB}

To demonstrate the behavior of the proposed procedure under simplified settings, we perform a simulation study that compares single-dataset random-effects meta-analysis with the standard random-effects meta-analytic approach using the classical framework. In a fully-factorial design, we simulate $K$ = 3, 10, 30, 100, and 300 cross-analytical estimates that re-analyze a finding with a true regression coefficient $\beta$ = 0.0 or 0.3 and a between-analyst heterogeneity $\tau$ = 0.0 or 0.1. In each repetition, the data were generated as follows; first, 100 observations were generated from a linear model with a single standard normally distributed predictor with regression coefficient $\beta$, and standard normally distributed residuals. Second, a linear regression was fitted to the simulated data, and the estimated regression coefficient and its associated standard error were saved. In conditions with no between-analyst heterogeneity ($\tau = 0$), the estimate was replicated $K$ times, simulating analysts performing the same analysis $K$ times. In conditions with between-analyst heterogeneity ($\tau = 0.1$), additional random deviations drawn from a normal distribution centered at zero with a standard deviation $\tau$ were added to the estimates, and the standard errors were multiplied by a factor drawn from uniform distribution from $0.75$ to $1.20$. The resulting estimates were synthesized using single-dataset random-effects (adjusted) and standard random-effects (unadjusted) meta-analysis implemented in the classical framework. We simulated each condition 10,000 times which reduced Monte Carlo standard errors (MCSEs) to a satisfactory level for all performance measures (maximum MCSE reported below). 

We implement the two-step single-dataset random-effects meta-analysis procedure with the restricted maximum likelihood estimator of the heterogeneity variance (REML) in both stages and equal weights, $1/K$, for all estimates, and the standard random-effects meta-analysis with REML using the \texttt{metafor} \texttt{R} package \citep{metafor}. Both methods converged in all but one repetition which was removed from the results. The methods are compared in terms of a) the comparison of the average standard error of the pooled estimates $\hat{\mu}$ (i.e., the mean of the standard errors returned by the method) and the empirical standard error of the pooled estimates $\hat{\mu}$ (i.e., the empirically observed standard deviation of meta-analytic estimates), b) the power and type-I error rate of the pooled estimate to reject the null hypothesis of no pooled effect using a level of 0.05 to threshold $p$-values, and c) the bias and root mean square error (RMSE) of the pooled estimate $\hat{\mu}$.\footnote{Performance for heterogeneity estimation is not compared because the $\tau$ estimate is the same for both methods, as the $\tau$ estimate for single-dataset meta-analysis is computed in the first stage.} The desired behavior in the simulation study is a) an average standard error of the pooled estimate corresponding to the empirical standard error of the estimate, b) a type-I error rate equal to 0.05, c) no bias and low RMSE. 

\begin{figure}
    \centering
    \includegraphics[width=0.75\linewidth]{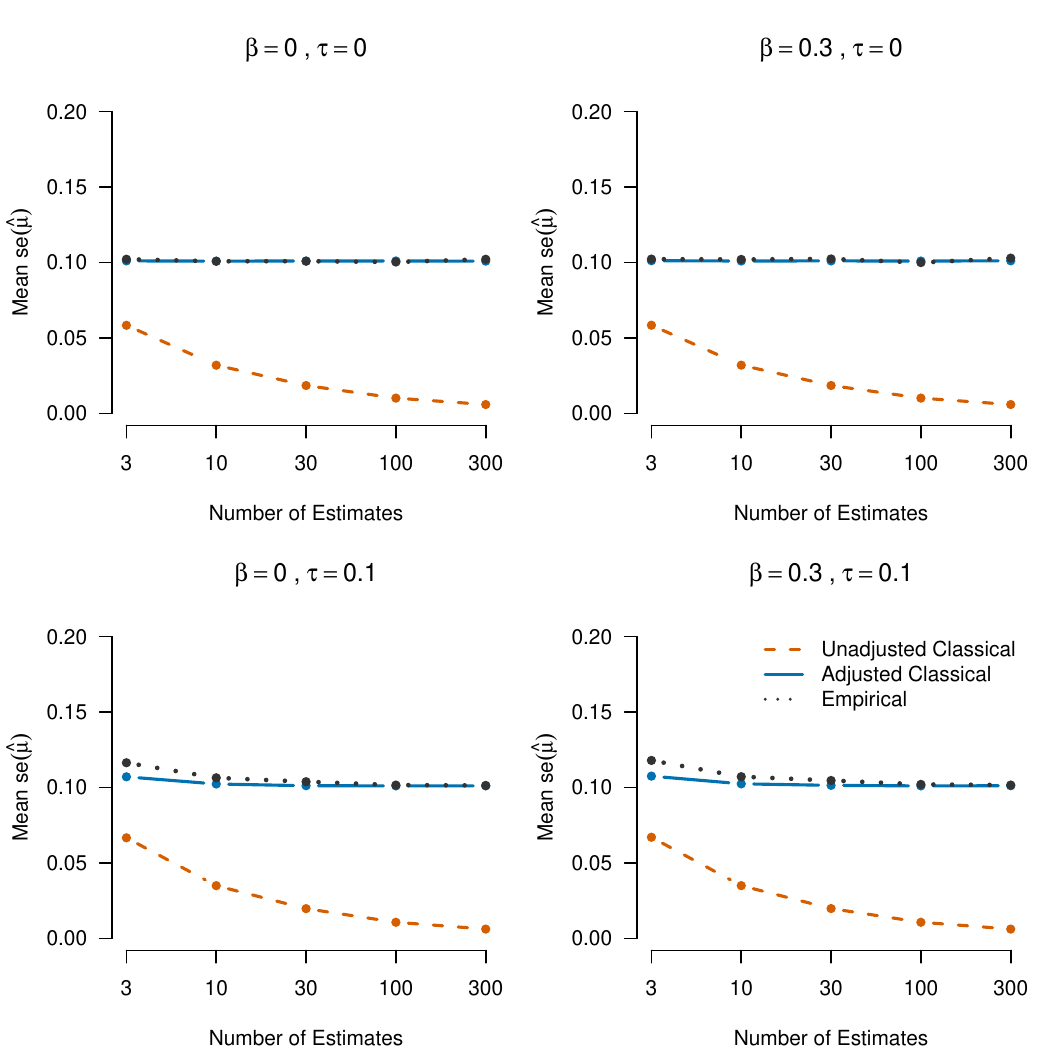}
    \caption{Comparison of the average standard errors of pooled cross-analytical estimates obtained from the standard random-effects meta-analysis (unadjusted; orange dashed line) and the single-dataset random-effects meta-analysis (adjusted; blue solid line) to the empirical standard error of the estimates (identical for both method, grey dotted line) across 3, 10, 30, 100, and 300 estimates. The top panels depict simulation conditions without between-analyst heterogeneity, and the bottom panels include heterogeneity. The left panels correspond to a true overall effect of 0, and the right panels to a true overall effect of 0.3. All MCSE are lower than 0.005.}
    \label{fig:sim}
\end{figure}

Figure~\ref{fig:sim} illustrates that the average standard error of single data-set random-effects meta-analysis (adjusted; blue solid line) tracks the empirical standard error of the estimate (black dotted line; the empirical standard error is practically identical for both methods) and the average standard error remains the same when increasing the number of estimates in the absence of between-analyst heterogeneity. In the presence of between-analyst heterogeneity, the average standard error tends to slightly underestimate the empirical standard error, most likely due to issues in estimating the between-analyst heterogeneity with a limited number of estimates, which is a common issue in meta-analysis \citep{valentine2017synthesizing, gonnermann2015no}. In contrast, the average standard error of standard random-effects meta-analysis (unadjusted; dashed orange line) decreases with increasing number of estimates and systematically underestimates the empirical standard error.

\begin{figure}
    \centering
    \includegraphics[width=0.75\linewidth]{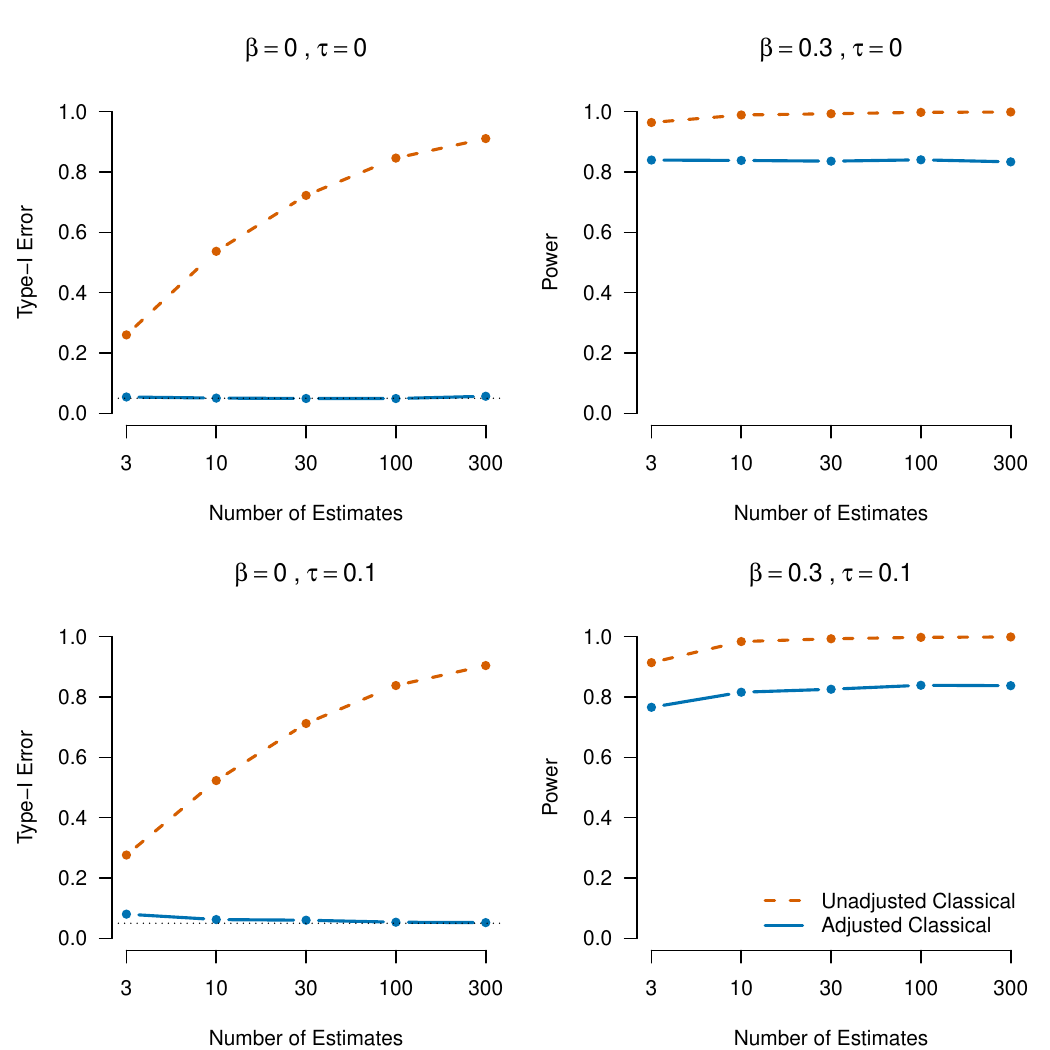}
    \caption{Comparison of the type-I error rate and power of the pooled cross-analytical estimates obtained from the standard random-effects meta-analysis (unadjusted, orange dashed line) and the single-dataset random-effects meta-analysis (adjusted, blue solid line) across 3, 10, 30, 100, and 300 estimates. The top panels depict simulation conditions without between-analyst heterogeneity, and the bottom panels include heterogeneity. The left panels correspond to a true overall effect of 0, and the right panels to a true overall effect of 0.3. All MCSE are lower than 0.002.}
    \label{fig:sim1}
\end{figure}

Figure~\ref{fig:sim1} illustrates the desired behavior of single-dataset random-effects meta-analysis in comparison to standard random-effects meta-analysis in all four conditions. While the type-I error is only slightly inflated when the number of estimates is small for the single-dataset meta-analysis, the type-I error systematically increases with the number of estimates for the standard random-effects meta-analysis. Due to its inflated type-I error, the power of the standard random-effects meta-analysis is higher.

\begin{figure}
    \centering
    \includegraphics[width=0.75\linewidth]{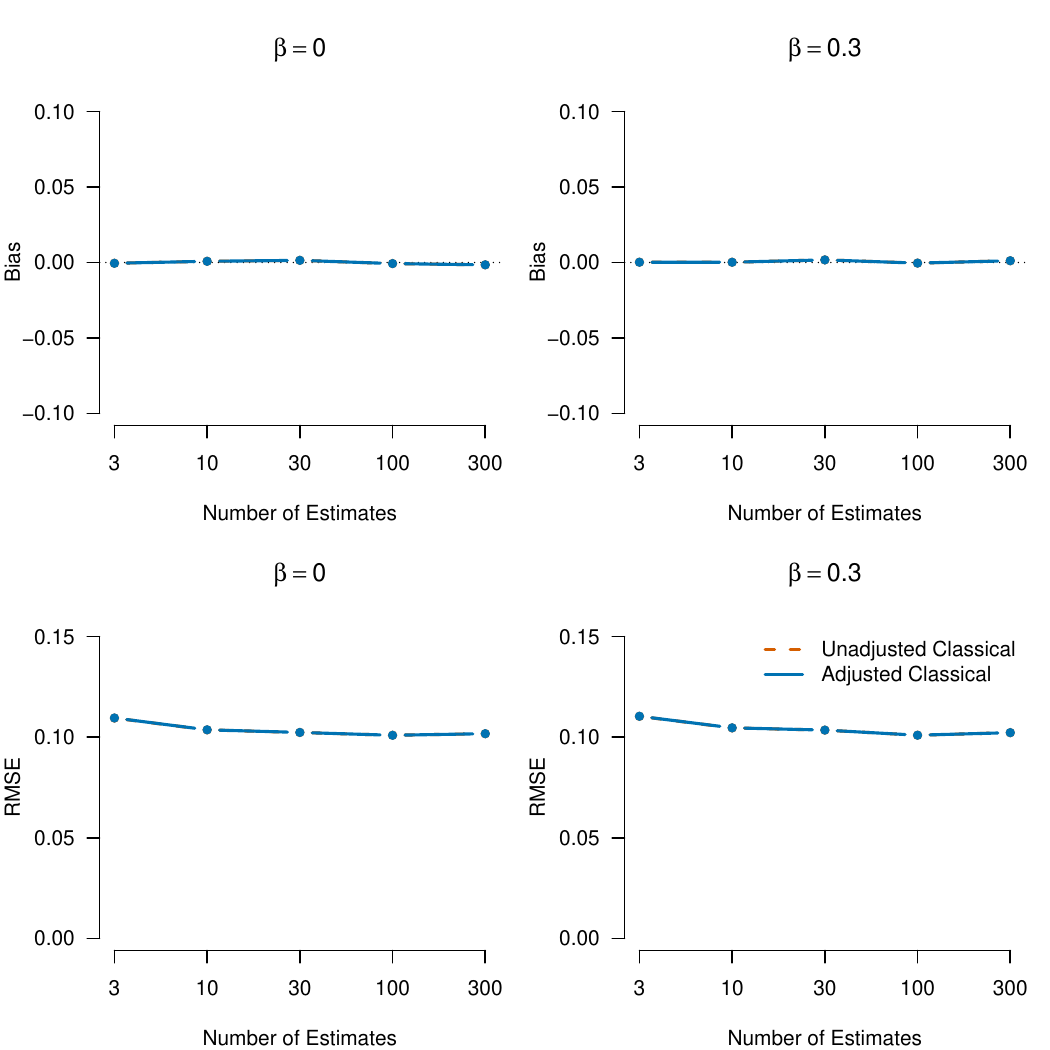}
    \caption{Comparison of the bias (left) and root mean square error (RMSE, right) of pooled cross-analytical estimates from the standard random-effects meta-analysis (unadjusted, orange dashed line) and the single-dataset random-effects meta-analysis (adjusted, blue, solid line) across 3, 10, 30, 100, and 300 estimates. Values are aggregated across both heterogeneity conditions. All MCSE are lower than 0.001.}
    \label{fig:sim2}
\end{figure}

Figure~\ref{fig:sim2} illustrates that the bias and RMSE of the meta-analytic estimates are virtually identical between the standard random-effects meta-analysis and the single-dataset random-effects meta-analysis across all examined conditions. This indicates that the proposed procedure appropriately adjusts only the uncertainty of the effect size estimate. The MC standard errors of the average standard error estimates are lower than 0.002 \citep{siepe2024simulation}. Code, computational environment, and other details are reported in a separate simulation study script available in our Open Science Framework project page.
\clearpage
\newpage

\bibliographystyle{apacite}
\bibliography{manuscript.bib,bib/all.bib,bib/software.bib,bib/unpublished.bib}

\end{document}